\documentclass[twocolumn, pra,showpacs,superscriptaddress]{revtex4}
\usepackage{graphicx}
\usepackage{dcolumn}
\usepackage{bm}
\usepackage{amsmath}

\begin{document}

\title{Ground State Properties of Anti-Ferromagnetic Spinor Bose gases in One Dimension}
\author{Yajiang Hao}
\email{haoyj@ustb.edu.cn}
\affiliation{Department of Physics, University of Science and Technology Beijing, Beijing 100083, China}
\date{\today }

\begin{abstract}
We investigate the ground state properties of anti-ferromagnetic spin-1 Bose gases in one dimensional harmonic potential from the weak repulsion regime to the strong repulsion regime. By diagonalizing the Hamiltonian in the Hilbert space composed of the lowest eigenstates of single particle and spin components, the ground state wavefunction and therefore the density distributions, magnetization distribution, one body density matrix, and momentum distribution for each components are obtained. It is shown that the spinor Bose gases of different magnetization exhibit the same total density profiles in the full interaction regime, which evolve from the single peak structure embodying the properties of Bose gases to the fermionized shell structure of spin-polarized fermions. But each components display different density profiles, and magnetic domains emerge in the strong interaction limit for $M=0.25$. In the strong interaction limit, one body density matrix and the momentum distributions exhibit the same behaviours as those of spin-polarized fermions. The fermionization of momentum distribution takes place, in contrast to the $\delta$-function-like distribution of single component Bose gases in the full interaction region.
\end{abstract}

\maketitle



\section{Introduction}
The spinor Bose-Einstein condensates (BECs) can be realized \cite{Ketterle98PRL,Champman2001PRL} as cold atoms are confined in an optical trap for the liberation of the atomic spin freedom degrees \cite{T.L.Ho,T.L.Ho2,Jap}. Its specific internal spin-mixing dynamics \cite{Law,Huang,He,WX Zhang} offer us a populated platform to investigate the spin-related interesting effects in conventional condensed matter physics. For example, spin domain \cite{spindomain,LZhou}, magnetic properties \cite{QGu,QGuPRB,AVinit,Higbie}, spin textures \cite{PhysRep,SYi}, and etc have been paid much attention \cite{RMP}. In addition the novel quantum phase \cite{Sadler,QGuPRL}, the effect of spin-orbit coupling \cite{HZhai,Lan}, and the spinor BECs of high spin also attracted \cite{Qi,Mithun,Gautam} many interests.

When quantum gases are confined in an anisotropic potential, the low dimensional quantum gas can be realized \cite{Paredes,Toshiya,Ketterle,Single1D}. For the strong correlation effect the quantum system in low dimension is one of the focuses of traditional condensed matter physics. The excellent controllability of cold atom system has stimulated significant interests in the low dimensional quantum gas \cite{RMP2011,RMP2012,RMP2013}. By controlling magnetic field the effective interaction strength can be tuned with the Feschbach resonance technique and confined induced resonance technique \cite{FR,CIR}. The quantum gas might evolve continuously from the weak interaction regime into the strong interaction regime.


In the weak interaction regime one-dimensional (1D) Bose gas exhibits single peak structure embodying the properties of Bosons, while in the strong interaction regime 1D Bose gas exhibits the same shell structure of multi-peaks as that of the spin-polarized fermions \cite{Hao2006}. The quantum mixtures with internal freedom degree of strong interaction can exhibit more rich properties such as composite fermionization \cite{Hao2009,Zollner2008,HaoCPL,Fang}, "spin-charge" separation \cite{TBC,Fush,Kleine},  and magnetic order \cite{GuanXW,Dehkharghani}, etc. Interestingly, although the density distribution of single-component Bose gas in strong interaction regime behave same as spin-polarized fermions in the coordinate space, momentum distribution still displays the $\delta$-function-like single peak structure in the low momentum region, which is the typical property of Bose gases.

It has been shown that a 1D spinor Bose gas displays phase separation and fermionization behaviour in the strong interaction limit \cite{DeuretzbacherPRL,HPu,Wang,HaoSpinor}. While with the increase of spin-dependent interaction, the fermionization will be weaken and phase separation disappears \cite{EPJD2016}. In this work we will investigate the ground state of 1D spinor Bose gas in different interacting regimes with the previously developed numerical diagonalization method \cite{Deuretzbacher,Hao2009,EPJD2016}. Particularly, the effect of interaction on one body density matrix and momentum distribution will be studied. We will diagonalize the Hamiltonian in the Hilbert space composed of the lowest eigenstates of single particle and spin components. Numerically the ground state wavefunction and therefore the interesting physical quanta of each components can be obtained. Compared with the previous work \cite{EPJD2016}, the evaluation of one body density matrix is time-consuming because we have to calculate the full one body density matrix rather than the density distribution, i.e., the diagonal part of one body density matrix. With the one body density matrix we can easily obtain the momentum distribution by its Fourier transformation. Besides the interaction effect, we will also investigate the spinor Bose gases of different magnetization. This can be implemented by diagonalizing Hamiltonian matrix in the subspace of Hilbert space with specific total spin.

The paper is organized as follows. In Sec. II, we briefly review the 1D spinor model in a harmonic trap and introduce numerical diagonalization method. In Sec. III, we present the density distributions, magnetization distribution, one body density matrix and momentum distribution for the 1D spinor Bose gas of anti-ferromagnetic spin-exchange interaction in the full repulsive interaction regime. The summary is given in Sec. IV.

\section{The model and method}

The spin-1 1D spinor Bose gas of two body contact interaction can be described by the second quantized Hamiltonian \cite{T.L.Ho,T.L.Ho2,Jap,Law}
\begin{eqnarray}
\hat{\mathcal{H}} &=&\int dx\left[ \hat{\Psi}_{\alpha }^{\dag }(x)(-\frac{
\hbar ^{2}}{2m}\frac{d^{2}}{dx^{2}}+V(x))\hat{\Psi}_{\alpha }(x)\right.  \nonumber \\
&&\left. +\frac{c_{0}}{2}\hat{\Psi}_{\alpha }^{\dag }(x)\hat{\Psi}_{\beta
}^{\dag }(x)\hat{\Psi}_{\beta }(x)\hat{\Psi}_{\alpha }(x)\right.  \\
&&\left. +\frac{c_{2}}{2}\hat{\Psi}_{\alpha }^{\dag }(x)\hat{\Psi}_{\alpha
\prime }^{\dag }(x)\mathbf{F}_{\alpha \beta }\cdot \mathbf{F}_{\alpha \prime
\beta \prime }\hat{\Psi}_{\beta \prime }(x)\hat{\Psi}_{\beta }(x)\right] \nonumber
\end{eqnarray}
with $m$ being atomic mass. $\hat{\Psi}_{\alpha}^{\dag}(x)$ [$\hat{\Psi}_{\alpha}(x)$] is the creation (annihilation) operator of $\alpha$-component atom at the position $x$, where $\alpha$=1, 0, and -1. Here $\mathbf{F}$ is the spin-1 Pauli matrix. The two body interaction consists of spin-independent $c_0$ term and spin-dependent $c_2$ term. The interaction constants $c_0$ and $c_2$ depend on the 1D effective interaction $g_f$ of the total spin-$f$ channel ($f=0,2$) as $c_0=\frac{g_0+2g_2}3$ and $c_2=\frac{g_2-g_0}3$, respectively. The 1D effective interaction constant $g_f$ can be tuned with Feshbach resonance technique and confined induced resonance, which is related to the external transvese confinement potential and the $s$-wave scattering length of the total spin-$f$ channel $a_f$ \cite{Olshanii,Dunjko,Petrov,HaoSpinor}
\begin{equation*}
g_f=\frac{4\hbar^2a_f}{ma_{\bot}^2}(1-\mathcal{C}\frac{a_f}{a_{\bot}})^{-1},
\end{equation*}
where $\mathcal{C}=1.4603$, and $a_{\bot}=\sqrt{\hbar/m\omega_{\bot}}$ with transverse trapping frequency $\omega_{\bot}$. By controlling magnetic field we can continuously tune the effective one-dimensional interaction from the weakly interacting regime to strongly interacting regime. Usually, the $s$-wave scattering lengthes of spin-$0$ channel and spin-$2$ channel are close to each other, so that the $c_2$ is greatly smaller than $c_0$. In the present paper, we will consider the case of $c_0=100c_2$. $c_2$ might be positive (anti-ferromagnetic interaction) or negative (ferromagnetic interaction). We will focus on the anti-ferromagnetic interaction.

For the quantum gas confined in a harmonic potential $V(x)=\frac12m\omega^2x^2$, it is natural to expand the field operator $\hat{\Psi}_{\alpha}(x)$ with the single particle wavefunctions of a particle in a harmonic trap (orbitals), i.e.,
\begin{equation*}
\hat{\Psi}_{\alpha }(x)=\sum_{i=1}^L\phi _{i}\left( x\right) \hat{b}_{i\alpha }.
\end{equation*}
Here the single particle wavefunction $\phi _i(x)=\frac {H_i(x)\exp (-x^{2}/2)}{\pi ^{1/4}\sqrt{2^{i }i !}}$ with Hermite polynomial $H_i(x)=i !\sum_{k=0}^{[i /2]}(-1)^{k}(2x)^{i -2k}/k!/(i -2k)!$ and the annihilation operator $\hat{b}_{i\alpha }$ annihilates one $\alpha$-component atom in the $i$th orbital. Therefore in the Hilbert space composed of the lowest eigenstates of single particle and spin components, the many body Hamiltonian can be formulated as
\begin{eqnarray*}
H &=&\sum_{i,\alpha }\mu _{i}\hat{b}_{i\alpha }^{\dagger }\hat{b}_{i\alpha }+%
\frac{c_{0}}{2}\sum_{\alpha \beta }\sum_{ijkl}I_{ijkl}\hat{b}_{i\alpha
}^{\dagger }\hat{b}_{j\beta }^{\dagger }\hat{b}_{k\beta }\hat{b}_{l\alpha }
\\
&&+\frac{c_{2}}{2}\sum_{\alpha \beta ;\alpha ^{\prime }\beta ^{\prime
}}\sum_{ijkl}I_{ijkl}\hat{b}_{i\alpha }^{\dagger }\hat{b}_{j\alpha ^{\prime
}}^{\dagger }\left( \mathbf{F}\right) _{\alpha \beta }\cdot \left( \mathbf{F}%
\right) _{\alpha ^{\prime }\beta ^{\prime }}\hat{b}_{k\beta ^{\prime }}\hat{b%
}_{l\beta },
\end{eqnarray*}
where $\mu _{i}=\left( i+\frac{1}{2}\right) \hbar \omega $ and $I_{ijkl} =\int dx\phi _{i}\left( x\right) \phi _{j}\left( x\right)\phi _{k}\left( x\right) \phi _{l}\left( x\right)$. We can obtain the ground state wavefunction by diagonalizing the Hamiltonian matrix in the Hilbert space. Because the contact interaction conserve the magnetization of the spinor Bose gases $M=\int dx m(x)=\int dx [\rho_1(x)-\rho_{-1}(x)]$, we can diagonalize the Hamiltonian in the subspace of total spin being conserved.

With the ground state wavefunction $\left\vert GS\right\rangle$, the one body density matrix of $\alpha$-component can be evaluated by
\begin{eqnarray*}
\rho _{\alpha }\left( x,y\right)  &=&\left\langle GS\left\vert \hat{\Psi}%
_{\alpha }^{\dag }(x)\hat{\Psi}_{\alpha }(y)\right\vert GS\right\rangle  \\
&=&\sum_{ij}\phi _{i}\left( x\right) \phi _{j}\left( y\right) \left\langle
GS\left\vert \hat{b}_{i\alpha }^{\dagger }\hat{b}_{j\alpha }\right\vert
GS\right\rangle,
\end{eqnarray*}
which denote the probability to find the particles at positions $y$ and $x$ in two successive measurements, respectively. Its diagonal part is the density distribution of $\alpha$-component, i.e.,
\begin{eqnarray*}
\rho _{\alpha }\left( x\right) =\sum_{ij}\phi _{i}\left( x\right) \phi _{j}\left( x\right) \left\langle
GS\left\vert \hat{b}_{i\alpha }^{\dagger }\hat{b}_{j\alpha }\right\vert
GS\right\rangle,
\end{eqnarray*}
and the total density profile is $\rho_{\text{tot}}(x)=\sum_{\alpha}\rho_{\alpha}(x)$.
The momentum distribution is the Fourier transformation of the one body density matrix $\rho_{\alpha}(x,y)$
\begin{equation*}
n_{\alpha }\left( k\right) =\frac{1}{2\pi }\int dxdy\rho _{\alpha }\left(
x,y\right) e^{-ik(x-y)}.
\end{equation*}
In the evaluation we use the dimensionless interaction $U_i$ ($i=0,2$ and $U_i=c_i/a_0$ with the harmonic length $a_0=\sqrt{\hbar/m\omega}$) and the position $x$ take the unit of $a_0$. For simplicity the notation will be preserved same as defined before.

\section{Ground state properties of anti-ferromagnetic 1D Bose gas}

In this section we will show the density distribution, magnetization distribution, one body density matrix and momentum distribution of anti-ferromagnetic 1D Bose gas for different magnetization.

\subsection{Density distribution and magnetization distribution }

\begin{figure}[tbp]
\includegraphics[width=3.0in]{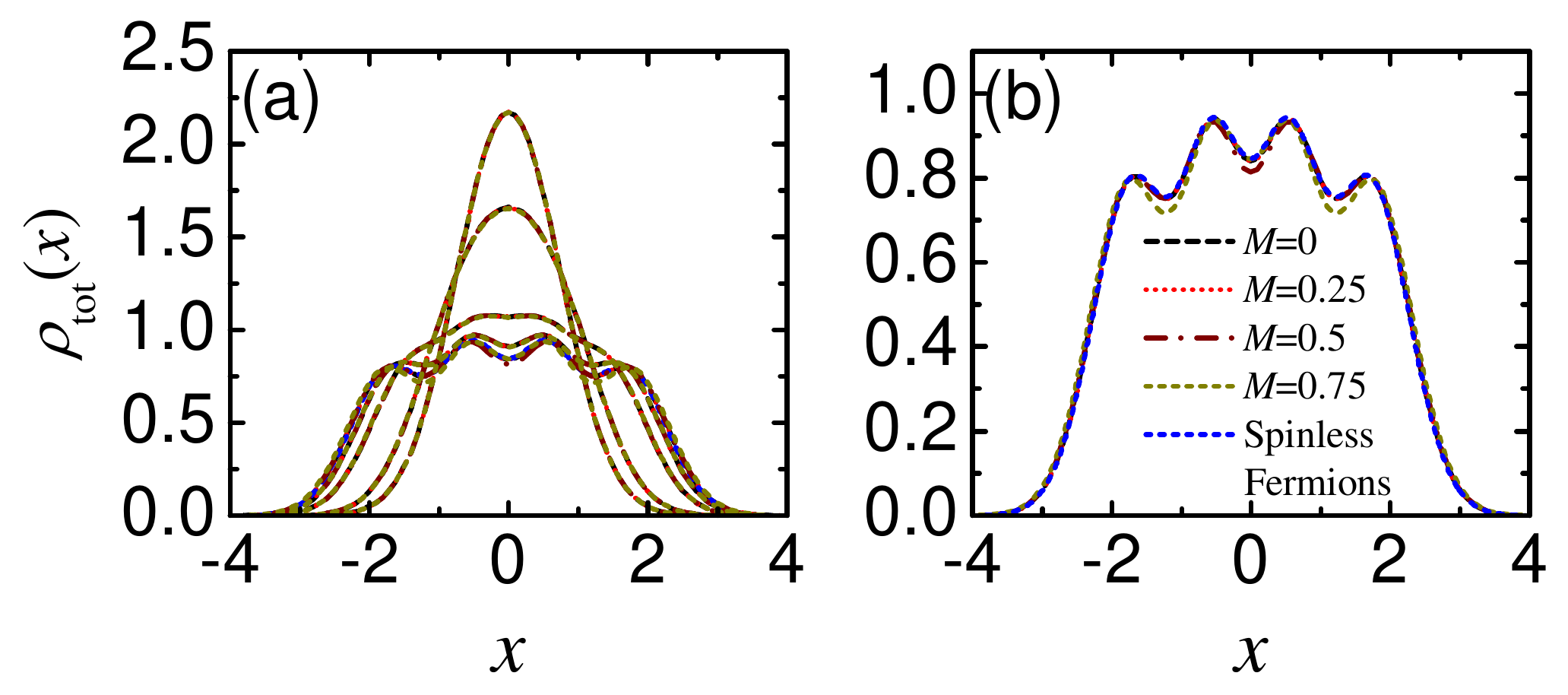}
\caption{(color online) Total density distribution of the ground state for $N=4$ for different magnetization $M$=0.0, 0.25, 0.5 and 0.75. Short dashed lines: spin-polarized Fermions. (a) From top to bottom: $U_0$=0.1, 1.0, 5.0, 10.0 and 50.0; (b) $U_0$=50.0. }
\end{figure}

In Fig. 1 the total density distribution of three components are displayed for magnetization $M$=0, 0.25, 0.5, and 0.75. It is shown that for different magnetization the total density profiles exhibit the same behaviors. In the weakly interacting regime, the density profiles show single peak structure embodying the property of Bose gases. As the atomic interaction becomes strong, Bose atoms distribute in wider regimes, and in the strongly interacting regime the Fermi-like shell structure are exhibited. The fermionized total density profiles are displayed in Fig. 1b for different magnetization. All of them match well with the density distribution of spinless fermions. The minor deviation results from the spin-dependent interaction, which affect the density distribution \cite{EPJD2016}.

\begin{figure}[tbp]
\includegraphics[width=3.0in]{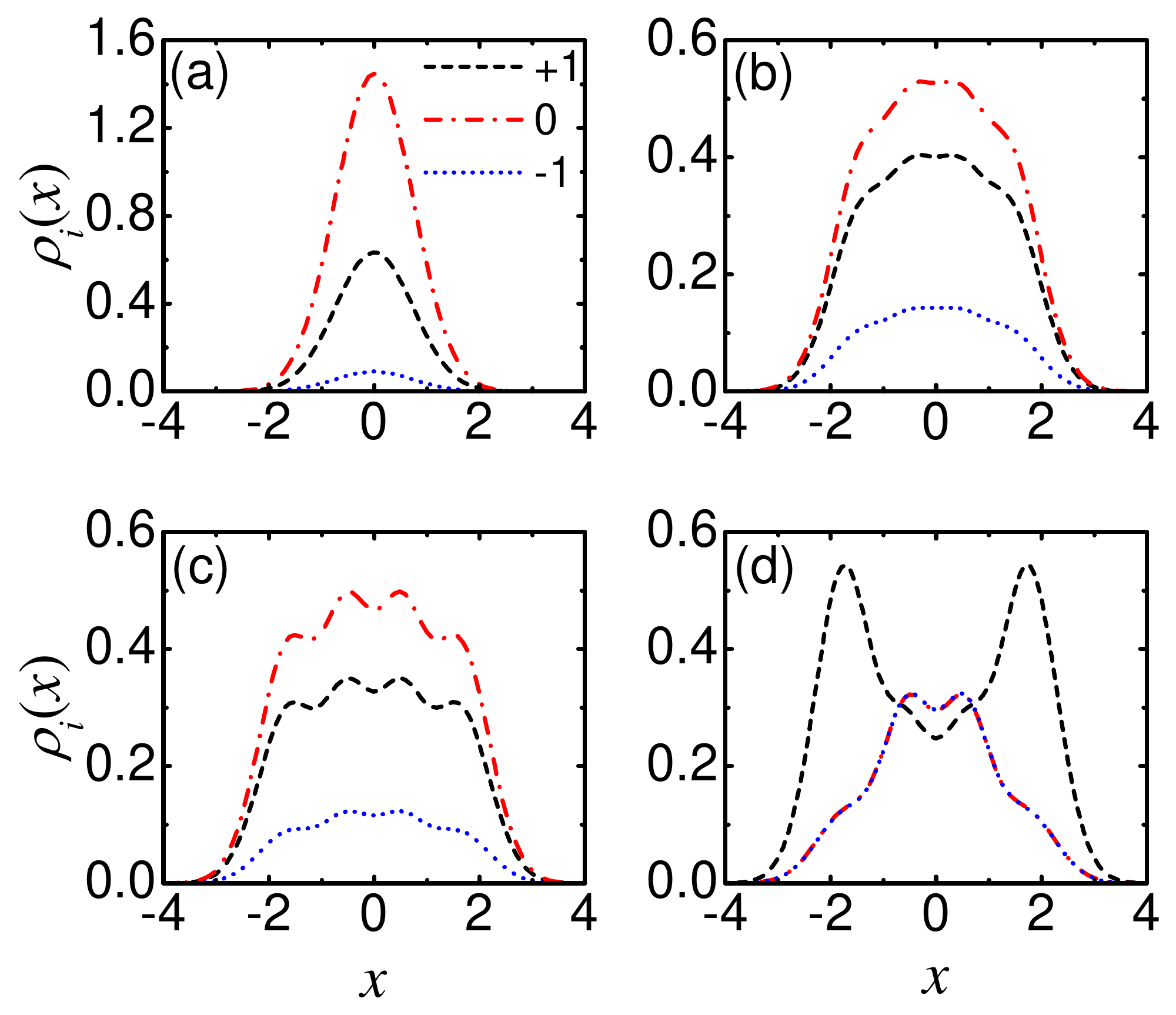}
\caption{(color online) The density distribution of each component for $N=4$ with $M=0.25$.(a) $U_0$=0.1; (b) $U_0$=5.0; (c) $U_0$=10.0; (d) $U_0$=50.0.}
\end{figure}
\begin{figure}[tbp]
\includegraphics[width=3.0in]{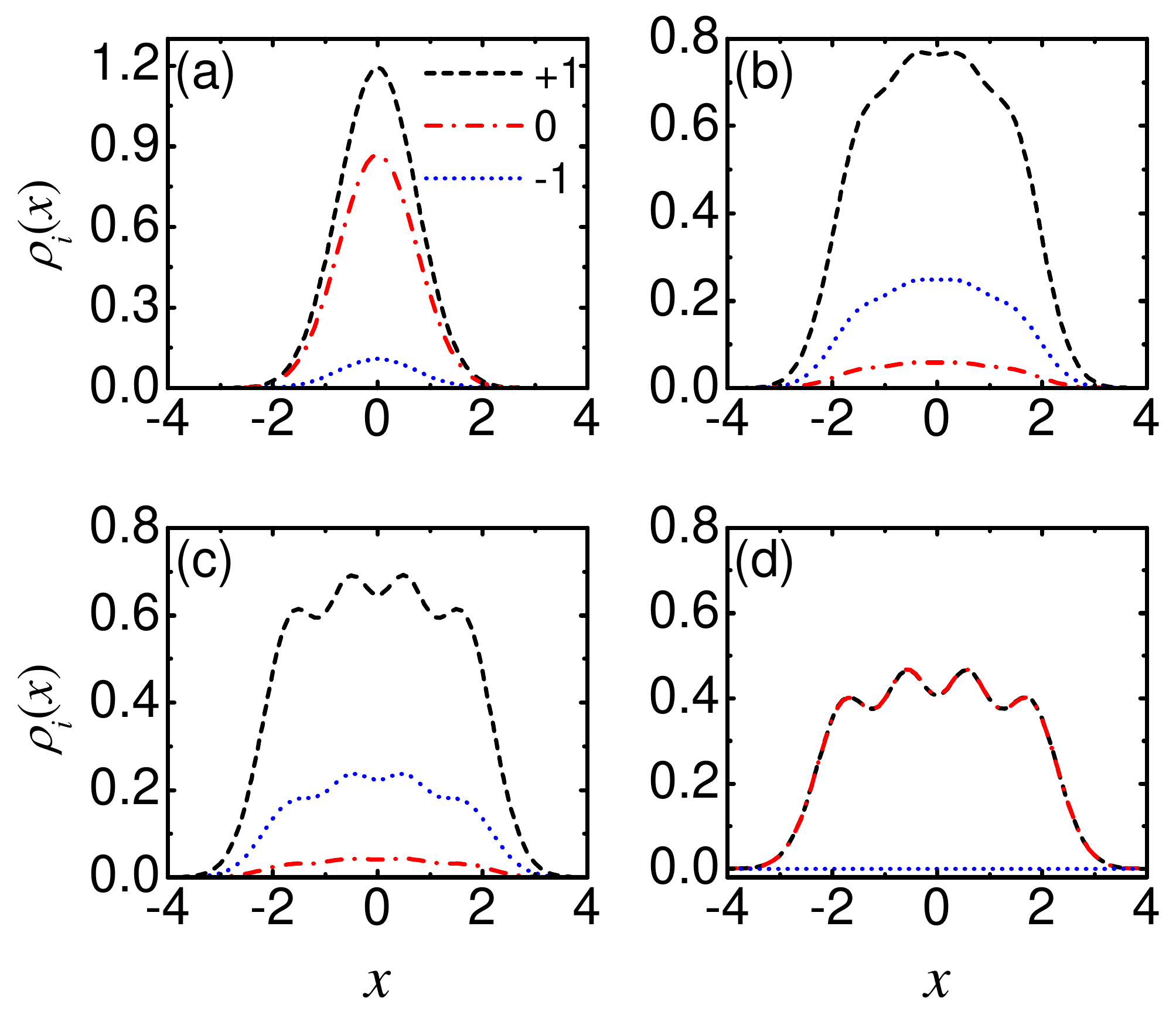}
\caption{(color online) The density distribution of each component for $N=4$ with $M=0.5$.(a) $U_0$=0.1; (b) $U_0$=5.0; (c) $U_0$=10.0; (d) $U_0$=50.0.}
\end{figure}

Although the 1D Bose gases with different magnetization exhibit the same total density distributions, each components behave differently, particularly in the strong interaction limit. The density distributions of each components for magnetization $M$=0.25 and 0.5 are displayed in Fig. 2 and in Fig. 3, respectively. In both situations, with the increase of interaction constants each components evolve from the single peak structure embodying Bose properties ($U_0=0.1$) to the Fermi-like shell structure ($U_0=10.0$) and each components exhibit the same fermionization behaviours. But the atom number of each components are different for $M=0.25$ and 0.5. In the former case the 0-components dominate for $U_0$=0.1, 5.0 and 10.0, while 1-component dominates in the latter case. In the strong interaction limit ($U_0=50.0$), the obvious distinctions are shown for different magnetization. For $M=0.25$ phase separation takes place, where those atoms in 1-component populate in the regime away from the trap center and other two components populate in the center regime of the trap with the exact same behaviour. For $M=0.5$ the (-1)-component disappear because of the strong atomic interaction and the spin-1 Bose gas become two-components of Bose gas composed of the components of 1-component and 0-component. Two components have same atom numbers and display the same fermionization behaviour.

\begin{figure}[tbp]
\includegraphics[width=3.0in]{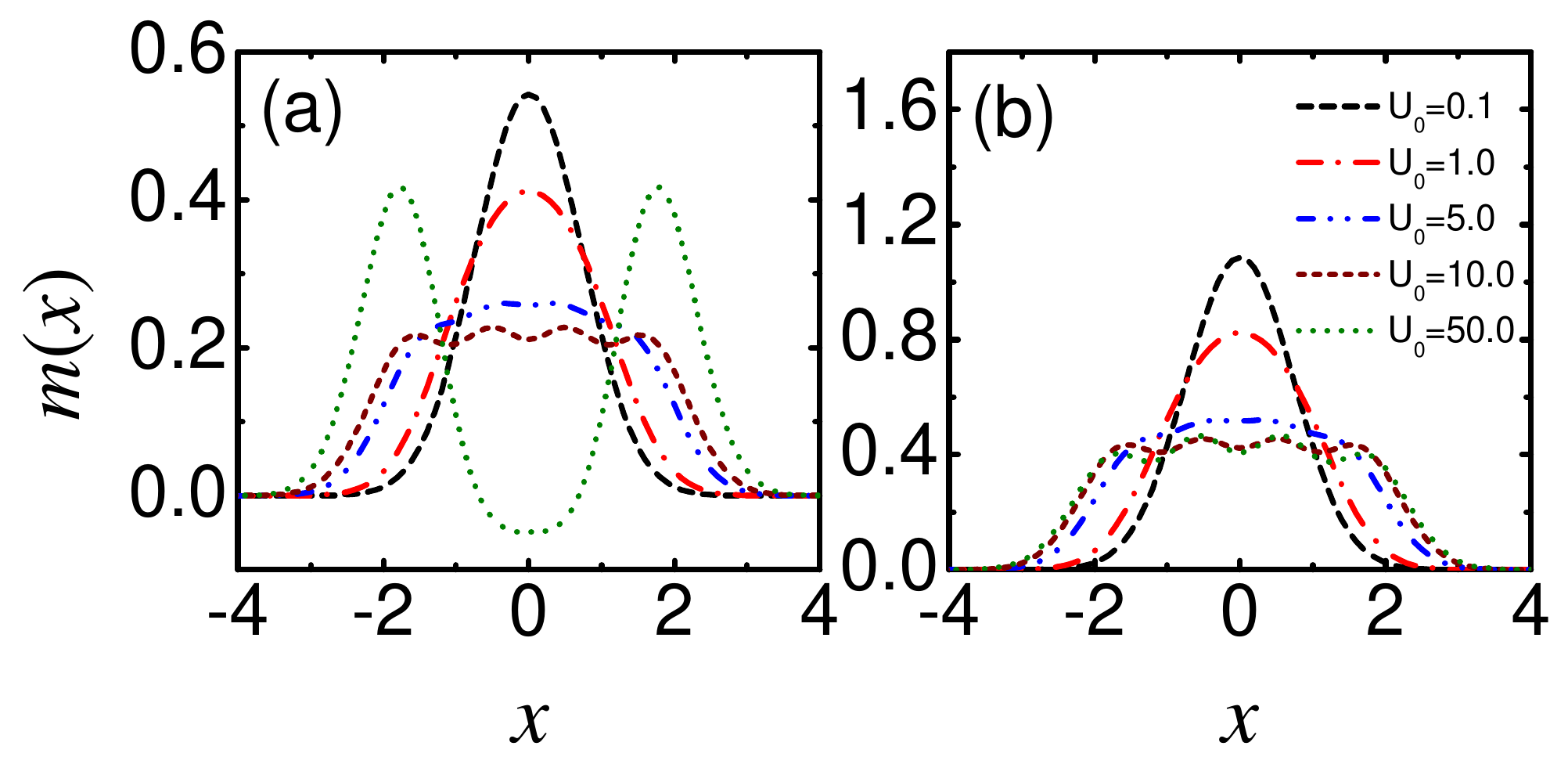}
\caption{(color online) Magnetization distribution of the ground state for different magnetization. (a) $M$=0.25; (b) $M$=0.5.}
\end{figure}

We plot the interaction effect of magnetization distribution $m(x)=\rho_1(x)-\rho_{-1}(x)$ for $M$=0.25 and 0.5 in Fig. 4. It is shown that in both cases the local magnetization change with the increase of interaction constant although the total magnetization are conserved. The magnetization distribution evolves from the single peak structure into the shell structure as the interaction becomes strong, which exhibits the fermionization behaviour in the strong interaction regime. The distinction for $M$=0.25 and 0.5 are displayed in the strong interaction limit ($U_0=50.0$), the latter show the fermionized shell structure but in the former case the magnetic domains emerge, where in the trap center regime the magnetization is negative and in the regime away from the trap center the magnetization is positive.

\subsection{One body density matrices of each components}

\begin{figure}[tbp]
\includegraphics[width=3.5in]{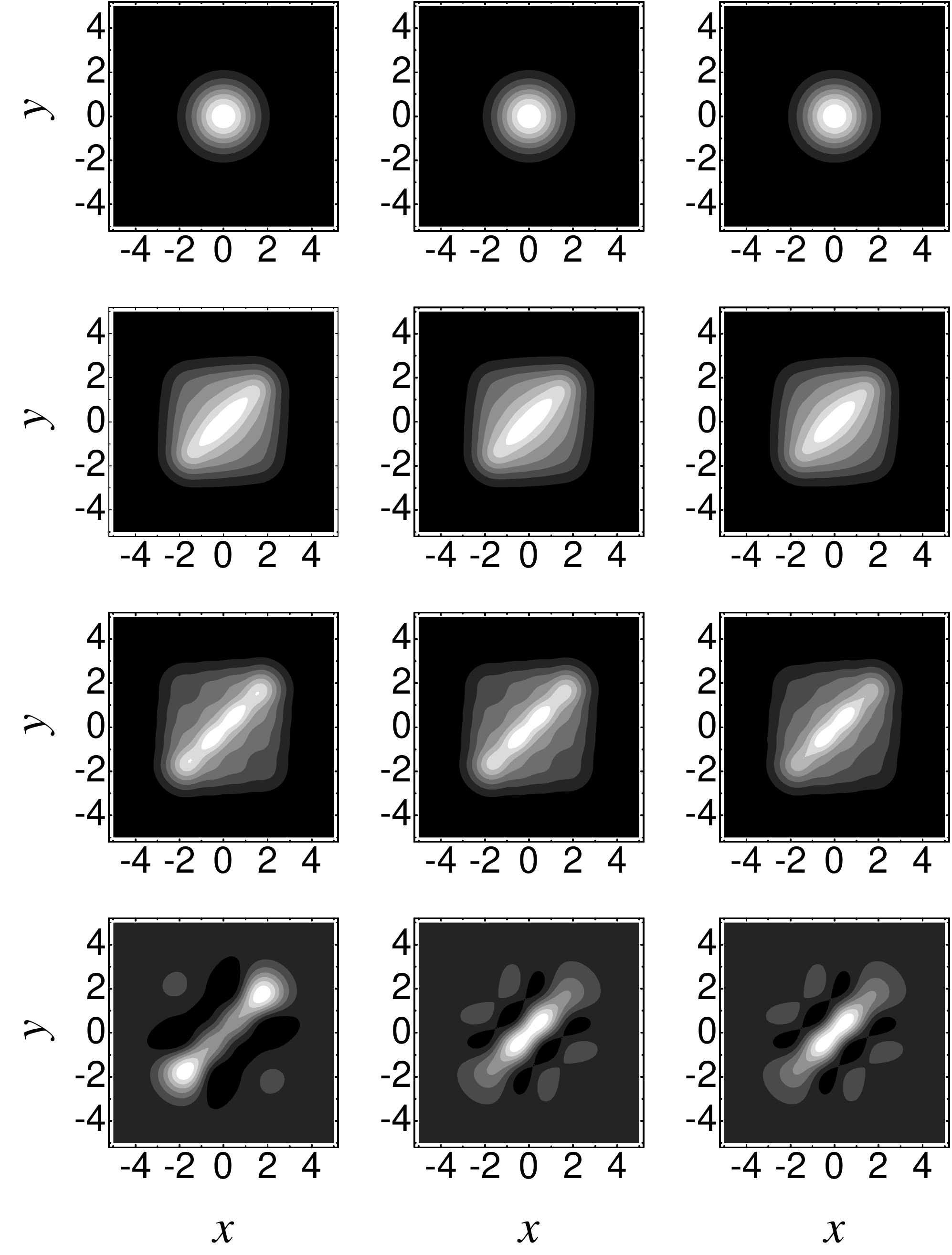}
\caption{The one body density matrix of the ground state for $N=4$ with $M=0.25$ for each components. Left column: 1-component; Middle column: 0-component; Right column: -1-component. $U_0$=0.1, 5, 10, and 50 from the first row to the last row. Brightness (dark) denotes the large (small) probability.}
\end{figure}

In Fig. 5 the one body density matrices of the ground state for each components are displayed for $M=0.25$. The diagonal elements $\rho(x,x)=\rho(x)$ denote density distribution, which has been shown in the previous subsection. Its off-diagonal elements relate to the off-diagonal long range order (ODLRO). It is shown that in the weak interaction regime and in the middle interaction regime three components display the same behaviours. Although the one body density matrices are diagonal dominant, the off-diagonal elements are not negligible. There exist ODLRO for each components. While in the strong interaction limit, one body density matrices become diagonal dominant and the off-diagonal elements are negligibly small such that the ODLRO disappear, which are similar to the spin-polarized Fermi gases. The diagonal regime of one body density matrix of 1-component exhibit properties different from other two components. 0-component and (-1)-component still exhibit the same behaviours. The atoms in 1-component prefer to appear in the region away from the trap center, while the atoms in other two components appear in the trap center region in more probabilities.

\begin{figure}[tbp]
\includegraphics[width=2.8in]{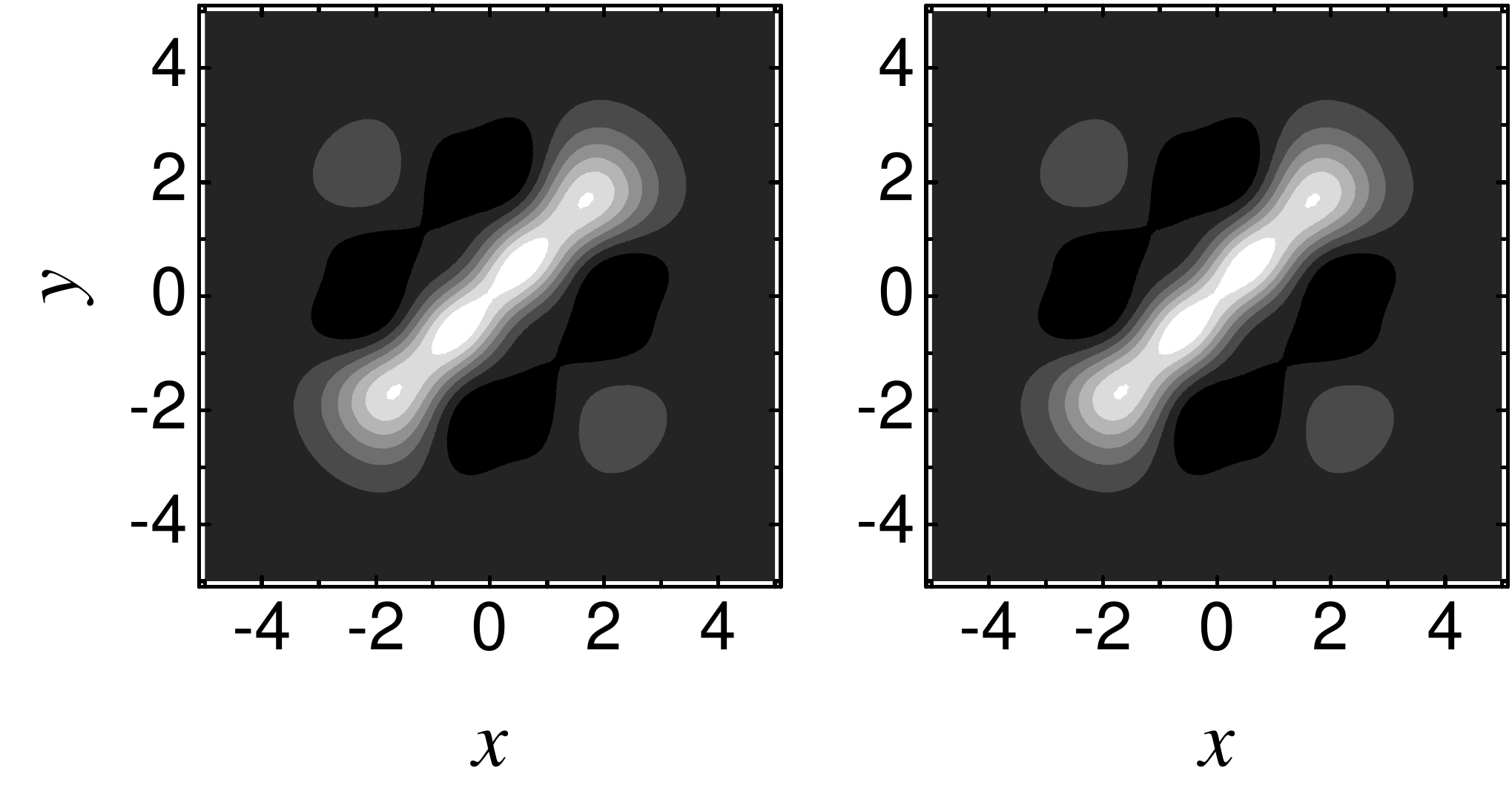}
\caption{The one body density matrix of the ground state for $N=4$ with $U_0$=50 and $M=0.5$ for 1-component (left column) and 0-component (right column).  Brightness (dark) denotes the large (small) probability.}
\end{figure}

In Fig. 6 we display the one body density matrices of ground state for 1-component and 0-component for $M=0.5$ and $U_0$=50. In this case, the (-1)-component disappear completely for the strong interaction.  For both 1-components and 0-component one body density matrices are diagonal dominant and the off-diagonal elements are negligibly small, which are same as the case of $M=0.25$. But for magnetization $M=0.5$, the 1-component and 0-component display the same properties that are similar to the spin-polarized Fermi gas of $N=4$. This is also shown in the density distribution.

\subsection{Fermionization of momentum distribution}

\begin{figure}[tbp]
\includegraphics[width=3.0in]{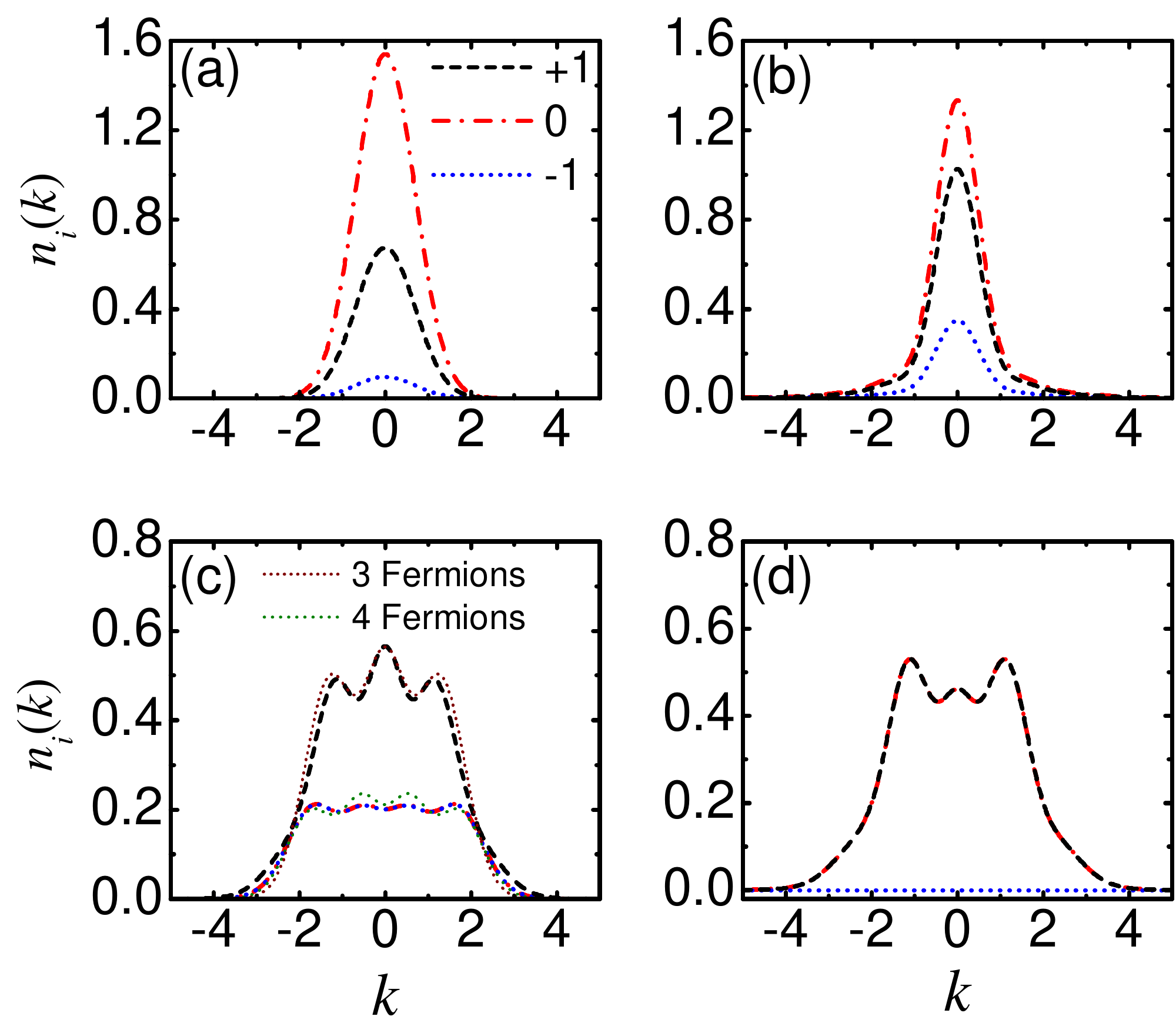}
\caption{(color online) Momentum distribution of the ground state for $N=4$ for each components. (a) $U_0$=0.1 and $M$=0.25; (b) $U_0$=5.0 and $M$=0.25; (c) $U_0$=50.0 and $M$=0.25; (d) $U_0$=50 and $M$=0.5.}
\end{figure}

The ground state momentum distribution of each components for $N$=4 and $M=0.25$ are displayed in Fig. 7 (a)-(c). It is shown that in the weak interaction regime ($U_0=0.1$) and in the middle interaction regime ($U_0=5.0$), each components show the sharp $\delta$-function-like single peak structure. Bose atoms populate in the regime of low momentum in large probability, and with the increase of momentum the population probabilities decrease rapidly. In the strong interaction limit ($U_0=50$), the momentum distribution of each components are fermionized. The momentum distribution of 1-component displays the shell structure of three peaks, and other two components display the shell structure of four peaks. Those are the properties of spin-polarized Fermi gases of three fermions and four fermions, respectively. As comparisons, the momentum distributions of spin-polarized fermions are plotted in the figures. For comparison, the fermion numbers are normalized to the atom number of each components. It is shown that the momentum distribution of each components of 1D spinor Bose gases behave completely same as those of spin-polarized fermions.

The momentum distributions of spinor Bose gases for $U_0=50$ and $M=0.5$ are displayed in Fig. 7 (d). 1-component and 0-component also exhibit the fermionized properties although their momentum distribution are not exact multi-peak shell structures.

\section{Summary}

In conclusion, we have investigated the ground state properties of anti-ferromagnetic 1D Bose gases of spin-1 from the weakly repulsive interaction regime to the strongly repulsive interaction regime. By numerically diagonalizing the Hamiltonian in the Hilbert space composed of the lowest eigenstates of single particle and spin components, we obtain the ground state wavefunction. Then the density distribution, one body density matrix and momentum distribution were evaluated in the full interacting regime.

It is shown that for different magnetization the total density distribution of three components exhibit the same fermionization behaviours with the increase of interaction. In spite of this, the density distributions of each components are different. Besides the atom numbers of each components, the main differences are shown in the strong interaction limit. For $M=0.25$, the phase separation takes places, while for $M=0.5$, the (-1)-component disappear and spinor Bose gas becomes two components of Bose gases, both of which display the same density profiles. The magnetization effect is also shown in the magnetization distribution in the strong interaction limit. For $M=0.25$ magnetic domains emerge at $U_0=50$, while for $M=0.5$ magnetization distribution exhibit fermionization shell structure.

In the weak interaction regime and middle interaction regime, the evaluation of one body density matrix shows that ODLRO exists in all of three components. In the strong interaction limit, the one body density matrix behave same as that of spin-polarized Fermi gas, which are diagonal dominant and the off-diagonal elements are negligibly small. This is consistent with the properties that the momentum distributions exhibit. In the strong interaction limit, the momentum distribution display the fermionization behaviours. It is greatly different from that of single component Bose gas in the strong interaction limit. The latter still exhibits the typical $\delta$-function-like momentum distribution embodying the properties of Bosons.

\begin{acknowledgments}
Y Hao thank Frank Deuretzbacher for useful discussion. The authors acknowledge the NSF of China (Grant No. 11004007) and ``the Fundamental Research Funds for the Central Universities".
\end{acknowledgments}


\end{document}